\documentclass[]{pasj01}
\Received{29-Mar-2019}
\Accepted{12-Apr-2020}
\Published{$\langle$publication date$\rangle$}
\SetRunningHead{Akahori et al.}{Discovery of Jets in Phoenix Cluster}

\begin{document}

\title{Discovery of Radio Jets in Phoenix Galaxy Cluster Center}
\author{
Takuya \textsc{Akahori}\altaffilmark{1},
Tetsu \textsc{Kitayama}\altaffilmark{2},
Shutaro \textsc{Ueda}\altaffilmark{3},
Takuma \textsc{Izumi}\altaffilmark{4},
Kianhong \textsc{Lee}\altaffilmark{5},
Ryohei \textsc{Kawabe}\altaffilmark{4,6,7},
Kotaro \textsc{Kohno}\altaffilmark{5,8},
Masamune \textsc{Oguri}$^{}$\altaffilmark{8,9,10}, and 
Motokazu \textsc{Takizawa}\altaffilmark{11}
}

\altaffiltext{}{
$^1$Mizusawa VLBI Observatory, National Astronomical Observatory Japan, 2-21-1 Osawa, Mitaka, Tokyo 181-8588, Japan \\
$^2$Department of Physics, Toho University, 2-2-1 Miyama, Funabashi, Chiba 274-8510, Japan \\
$^3$Academia Sinica Institute of Astronomy and Astrophysics (ASIAA), No. 1, Section 4, Roosevelt Road, Taipei 10617, Taiwan \\
$^4$National Astronomical Observatory Japan, 2-21-1 Osawa, Mitaka, Tokyo 181-8588, Japan \\
$^5$Institute of Astronomy, The University of Tokyo, 2-21-1 Osawa, Mitaka, Tokyo 181-0015, Japan\\
$^6$The Graduate University for Advanced Studies (SOKENDAI), 2-21-1 Osawa, Mitaka, Tokyo 181-8588, Japan \\
$^7$Department of Astronomy, The University of Tokyo, 7-3-1 Hongo, Bunkyo, Tokyo 113-0033, Japan \\
$^8$Research Center for the Early Universe, The University of Tokyo, 7-3-1 Hongo, Bunkyo-ku, Tokyo 113-0033, Japan \\
$^{9}$Department of Physics, The University of Tokyo, 7-3-1 Hongo, Bunkyo-ku, Tokyo 113-0033, Japan \\
$^{10}$Kavli Institute for the Physics and Mathematics of the Universe (Kavli IPMU, WPI), University of Tokyo, Kashiwa, Chiba 277-8583, Japan \\
$^{11}$ Department of Physics, Yamagata University, 1-4-12 Kojirakawa-machi, Yamagata, Yamagata 990-8560, Japan
}

\email{takuya.akahori@nao.ac.jp}


\maketitle

\begin{abstract}
We report the results of the Australia Telescope Compact Array (ATCA) 15 mm observation of the Phoenix galaxy cluster possessing an extreme star-burst brightest cluster galaxy (BCG) at the cluster center. We spatially resolved radio emission around the BCG, and found diffuse bipolar and bar-shape structures extending from the active galactic nucleus (AGN) of the BCG. They are likely radio jets/lobes, whose sizes are $\sim 10$--$20$~kpc and locations are aligned with X-ray cavities. If we assume that the radio jets/lobes expand with the sound velocity, their ages are estimated to be $\sim 10$~Myr. We also found compact radio emissions near the center and suggest that they are another young bipolar jets with $\sim 1$~Myr of age. Moreover, we found extended radio emission surrounding the AGN and discussed the possibility that the component is a product of the cooling flow, by considering synchrotron radiation partially absorbed by molecular clumps, free-free emission from the warm ionized gas, and the spinning dust emission from dusty circum-galactic medium.
\end{abstract}

\section{Introduction}
\label{section1}

A galaxy cluster is permeated with the hot intra-cluster medium (ICM) emitting electron thermal bremsstrahlung and metal line-emission in X-ray. If only the loss of thermal energy by the X-ray radiation is realized, it leads the quasi-hydrostatic evolution of the ICM under the gravitational potential \citep{mas04, aka06}, and results in a catastrophic cooling-flow of a huge amount of cooled ICM at the cluster center (see \cite{fab94} for a review). X-ray observations, however, have found no such massive cooled gas at the centers of nearby clusters (e.g., \cite{tam01, mak01, pet01, lew02}). Instead, many clusters indicate the bottom of the ICM temperature called ``cool-core" (e.g., \cite{san09}). The core radius, within which a radial profile of the ICM density has a flat slope, of a cool-core cluster tends to be one third of those of other relaxed clusters \citep{om02, om04}.

Cool-core clusters tend to possess cD galaxies or central brightest cluster galaxies (BCGs) (e.g. \cite{aka05}), which often exhibit active galactic nucleus (AGN) jets. It has been considered that energy injection by the jets suppresses runaway cooling of the ICM and forms its cool-core profile, known as the AGN feedback scenario (e.g., \cite{rus04}; \cite{fab12} for a review). Massive star-burst activities expected under the classical cooling-flow should also be suppressed in central galaxies because of less inflow of cooled gas. Actually, BCGs are giant elliptical galaxies which are inactive star-forming systems in general, although minor star-forming activities can be seen in the local Universe; NGC1275 in the Perseus cluster hosts blue, young star clusters \citep{hol92, con01} and star-forming regions \citep{can14}.

The Phoenix galaxy cluster (SPT-CLJ2344-4243) seems to be an exception from the picture established in the local Universe. The Phoenix cluster is a distant cluster located at the redshift of $z=0.596$. It is a very massive cluster with the mass of $2.5\times 10^{15}$~M$_\odot$ within the virial radius \citep{mcd12}. One of the spectacular features of this cluster is that there is the BCG possessing an unprecedented star-burst of $800$~${\rm M_\odot/yr}$ \citep{mcd12, mcd13}. Moreover, it is suggested that the BCG possesses an ultra-heavy ($1.8 \times 10^{10}$~${\rm M_\odot}$) and fast-growth ($60$~${\rm M_\odot/yr}$) super-massive black hole (SMBH) \citep{mcd13}. Around the SMBH, huge cold molecular gas of $2.1 \times 10^{10}$~M$_\odot$ is expected according to the intensity of carbon monoxide line, CO(3-2) \citep{rus17}. Based on the X-ray luminosity and its intrinsic absorption \citep{ued13}, this AGN is classified into the type-2 quasar, which is quite rare in BCGs (only two samples known to date). These features imply that the massive cold-gas inflow expected in the classical cooling-flow scenario is realized in the Phoenix cluster (\cite{mcd12}; 2013; 2014; 2015). 

While the Phoenix cluster center exhibits some indirect evidence of classical cooling-flow, it is notable that the surrounding environment is rather similar to those in nearby cool-core clusters. For example, global X-ray morphology is symmetric and round, so that it is unlikely to consider that a major merger of clusters is taking place. \citet{ued13} found that global ICM temperature structure bottoms to 3~keV from 11~keV of outer regions, implying that the Phoenix cluster has a cool core. Moreover, \citet{mcd15} reported bipolar X-ray cavities, which are often seen along with AGN activities in several nearby cool-core clusters including the Perseus cluster \citep{fab11}. Therefore, AGN feedback seems to be taking place in the Phoenix cluster, without suppressing the massive star-formation and the rapid black-hole growth. 

Taking a close look at this cluster will therefore give us unique understanding for co-evolution of the ICM, the BCG, and the SMBH. Key information missing so far is whether or not there are radio jets/lobes, which are firm evidence of AGN feedback activities. \citet{hog15a} and \citet{hog15b} presented comprehensive studies of the radio properties and variabilities of BCGs. A relatively-steep spectrum would be evidence of radio jets/lobes, while AGN core emission indicates a flatter spectrum particularly at high frequencies. A more straightforward way to find jets/lobes is to spatially resolve potential radio emissions. \citet{van14} reported that there is diffuse radio emission in the Phoenix cluster. But radio jets/lobes have never been resolved. There was only one long-baseline radio observation ever. That is a 16 cm observation of the Australia Telescope Compact Array (ATCA) with the longest, 6 km array configuration (Project C2585). The authors reduced the data, however, no structure was resolved even with the finest beam size of $6 \farcs 92 \times 3 \farcs 84$ at the highest band of 3.1~GHz of the observation.

In this paper, we report the first sub-arcsec resolution centimeter observation of the Phoenix cluster center. We chose the 15~mm band, which has never been performed for this cluster and is the best suited to our objective because it is possible to perform high-resolution observation. We resolved emissions from AGN core and radio lobes successfully. This paper is organized as follows. We summarize our observation and data reduction in Sections 2 and 3, respectively. Results are shown in Section 4, where we combine the data of {\it Atacama Large Millimeter/submillimeter Array} ({\it ALMA}) Band 3. Our discussion is presented in Section 5, referring the archival data of radio, optical, and X-ray observations so as to extend our discussion for the origin of the radio emissions we detected with ATCA. Our conclusion is made in Section 6. Unless otherwise specified, we consider the standard $\Lambda$CDM cosmology model with the cosmological parameters, $H_0=70$~km~s$^{-1}$~Mpc$^{-1}$, $\Omega_{\rm M}=0.3$, and $\Omega_{\rm \Lambda}=0.7$, leading to $1'' \sim 6.7$ kpc at the source redshift of $z=0.596$.

\section{Observation and Data Reduction}
\label{section2}

\begin{table}
\tbl{Summary of observing specifications of ATCA.}{%
\begin{tabular}{lc}
\hline
Pointing (right ascension) & 23:44:44 \\
Pointing (declination) & -42:43:15 \\
Date and time 1 (UT) & 2017 November 1 06:30 -- 16:30 \\
Date and time 2 (UT) & 2017 November 2 06:30 -- 16:30 \\
Maximum baseline & 6~km \\
Center frequency$^\dagger$ (MHz) & 17000, 19000 \\
Bandwidth$^\dagger$ (MHz) & 2048, 2048 \\
Setup calibrator & 2251+158 \\
Flux calibrator & 1934-638 \\
Bandpass calibrator & 1921-293 \\
Pointing/gain/phase calibrator & 2333-415 \\
Target on-source time (min) & 760\\
\hline
\end{tabular}}\label{t01}
\begin{tabnote}
$^\dagger$CABB recorded two IF bands.
\end{tabnote}
\end{table}

Our observations were made with the ATCA in November 1 and 2 in 2017 (Project C3190, see Table \ref{t01}). A total observing time of the target were 360 minutes in November 1 and 400 minutes in November 2. The 6A array configuration was adopted, where all six antennas were mounted in the east-west track with the 6~km baseline at the maximum. The 1M-0.5k Compact Array Broadband Backend (CABB, \cite{wil11}) receiver mode was selected to utilize the 2048 channels with a 1 MHz spectral window at the 15~mm band, and full spectro-polarimetric data were recorded. Two intermediate frequencies (IFs) were set at the center frequencies of 17000 MHz for IF1 and 19000 MHz for IF2. The fields of view was $2\farcm8$ ($2\farcm5$) for IF1 (IF2), and the largest well-imaged structure was $\sim 32''$ at 18 GHz (Section 1.7.1 of ATCA Users Guide).

In each day, we observed the initial setup calibrator 2251+158 for 30 minutes, the flux-scale calibrator 1934-638 for 6 minutes, the bandpass calibrator 1921-293 for 6 minutes, a repetitive sequence, and the gain/phase calibrator 2333-415 for 2 minutes in this order. The repetitive sequence included observations of the pointing calibrator 2333-415 for 2 minutes, the gain/phase calibrator 2333-415 for 2 minutes, and the target for 8 minutes, where 2333-415 is sufficiently bright (699 mJy at 15 mm) and close ($2^\circ.01$ from the target). The sequence was repeated nine times in November 1 and ten times in November 2. This long-time tracking provided us with good uv coverage. The system temperature was around 45 K (meridian passage) -- 82 K (setting) and those were stable during the observing days. 

Data reduction was performed with the MIRIAD software \citep{sau95} version 1.5 following the standard procedures unless otherwise specified. All data were loaded by the MIRIAD task \textsc{atlod} with options \textsc{birdie}, \textsc{xycorr}, \textsc{rfiflag}, \textsc{opcorr}, and \textsc{noauto}. Edge 40 channels (32 MHz bandwidth) were removed by \textsc{uvflag}. We performed \textsc{pgflag} iteratively until radio frequency interferences (RFIs) disappeared in visibility plots. We adopted a modern flagging recipe in which we looked at visibility variations of Stokes V at first (Section 4.3.5 of ATCA Users Guide). The band-pass solution was made by \textsc{mfcal} using 1921-293. The solution was transferred to 2333-415 by \textsc{gpcopy}, and then \textsc{gpcal} was carried out to obtain the gain and phase solutions, with options \textsc{xyvary} and \textsc{qusolve}. The solution was transferred to 1934-638, and \textsc{gpcal} was carried out to obtain the absolute flux scale, with options \textsc{xyvary}, \textsc{qusolve}, and \textsc{nopol}. The error of the absolute value was computed by bootstrap estimation using \textsc{gpboot}, from 1934-638 to 2333-415. Finally, all solutions were applied to the target data from 2333-415 using \textsc{gpcopy}.

The visibility was transformed into real-space images using \textsc{invert}, where the robustness parameter, $r$, were explored from 0.5 to 2.0 to check the trade-off between the angular resolution and the image rms noise. The beam full width at half maximum (FWHM) of major/minor axes and the beam position angle are $1 \farcs 03 \times 0 \farcs 46$ and 6.5 degree, respectively, at 18 GHz with $r=0.5$. The images were cleaned by multi-frequency CLEAN task \textsc{mfclean}, where we chose the best iteration number which minimized the rms noise in the cleaned image. Using the brightest compact source at the center of the image (Source C1, see the next section), we performed phase self-calibration three times with the solution intervals of 15 minutes, 5 minutes, and 1 minute, updating the CLEAN model. The self-calibration improved the signal-to-noise ratio (S/N) of Source C1 from 251 to 365.  We finally achieved the image rms noise level almost comparable to the theoretical value, and found no significant emission of Stokes Q and U in the image. The upper limits (the $1\sigma$ noise levels) are 0.00639~mJy/beam and 0.00640~mJy/beam for Stokes Q and U, respectively, with $r=2.0$.

\section{Results}
\label{section3}

\subsection{Synthesized Images}\label{section3.1}

\begin{figure*}[tp]
\begin{center}
\FigureFile(120mm,120mm){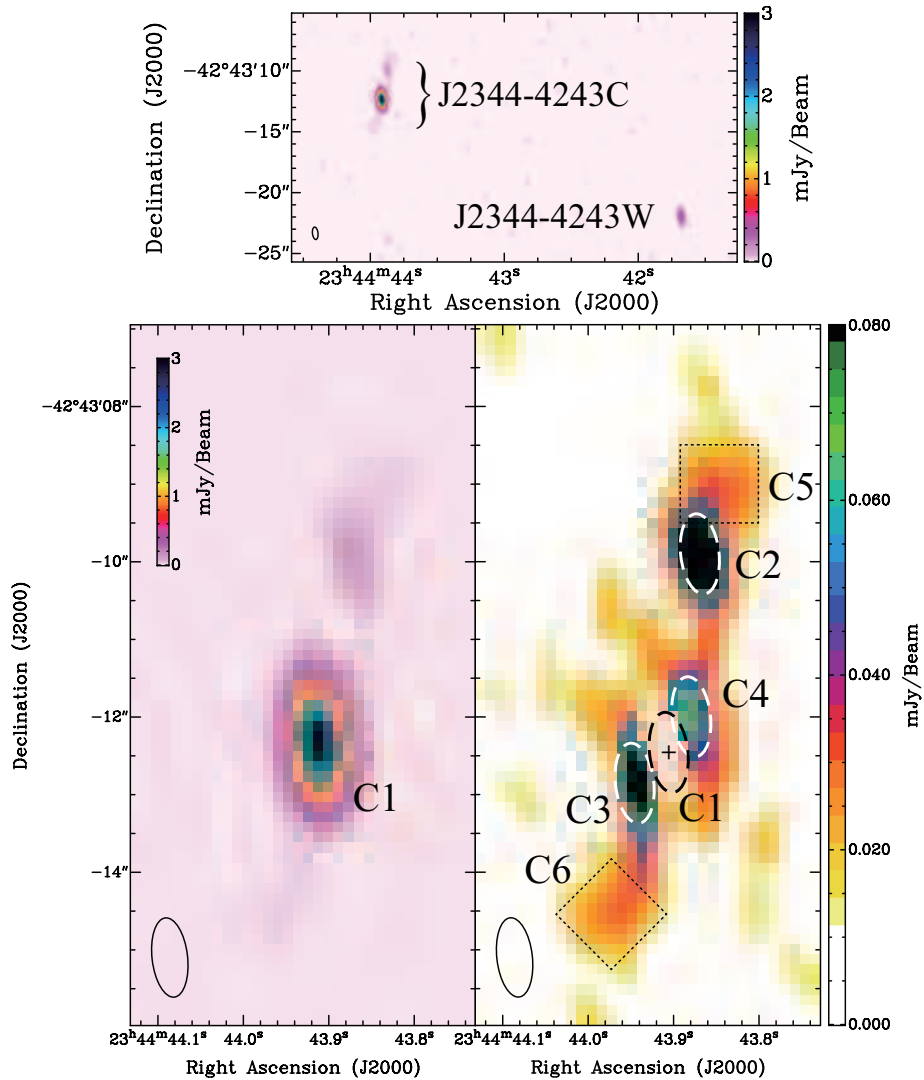}
\end{center}
\caption{
(Top) ATCA 18~GHz total intensities (Stokes I) map of the Phoenix cluster of galaxies. (Bottom left) An enlarged map around J2344-4243C. (Bottom right) Same as left but Source C1 (the black dashed eclipse) is subtracted. Also shown are Sources C2, C3, and C4 as the white dashed eclipses. The dotted boxes with $1''\times 1''$ indicate the northern (C5) and southern (C6) diffuse emissions (see the text). The image rms noise level ($1\sigma$) is 0.00612~mJy/beam. The left-bottom eclipse of each panel indicates the synthesized beam size. 
}
\label{f01}
\end{figure*}

The top panel of figure~\ref{f01} shows an image at 18 GHz, which corresponds to the rest-frame frequency of 28.7 GHz at z = 0.596, with a 4096 MHz bandwidth and $r=0.5$. We find two radio components. The brightest, extended component, J2344-4243C, is located at the center of the Phoenix cluster and is associated with the BCG. The other, compact component, J2344-4243W or Source W, is 26.25 arcsec away from the cluster center to the west direction. A zoom-in map of J2344-4243C is shown in the bottom left panel of figure~\ref{f01}. There is a compact source, J2344-4243C1 or Source C1 hereafter, at the center. 

We evaluated flux densities of Sources C1 and W using the MIRIAD task \textsc{imfit} with the option \textsc{object=point}. We obtained the flux densities of 3.395 $\pm$ 0.132~mJy for Source C1 and 0.478 $\pm$ 0.019~mJy for Source W both at 18~GHz. We subtracted Sources C1 and W from the image, and the resultant residual map is shown in the bottom right panel of figure~\ref{f01}. We clearly see some striking features in the residual map. First, there is a bright spot in the north of Source C1. We label the spot as Source C2. Second, we see two spots near Source C1. We label them as Sources C3 and C4. Third, including Sources C3 and C4, there are diffuse emission around the center. Finally, we see another two diffuse emissions in the north (labelled C5) and south (C6). They connect to central diffuse emission with straight bar-shape structures.

As with Sources C1 and W, Sources C2, C3, and C4 seem to be unresolved sources. Using the same MIRIAD task as Source C1, we evaluated the flux densities of 0.147 $\pm$ 0.026~mJy, 0.108 $\pm$ 0.010~mJy, and 0.077 $\pm$  0.015~mJy for Sources C2, C3, and C4, respectively, at 18~GHz. 

All of the features mentioned in this subsection are also seen in the image made with $r=2.0$. Those features are simply blurred, because of the worse angular resolution by 40~\% but with the better image rms noise level ($1\sigma$) of 0.00482~mJy/beam compared to 0.00612~mJy/beam with $r=0.5$. We tried various self-calibration procedures with different solution intervals for phase and gain, and confirmed that the labelled structures and the central extended emission remain. Therefore, we conclude that those features are all real.

With the total intensity and the upper limit of the polarized intensity, $P=\sqrt{Q^2+U^2}$, we got the upper limits of the polarization fraction, $P/I$. Those are 0.27~\%, 5.94~\%, 11.3~\%, 18.1~\%, and 1.89~\% of Sources C1, C2, C3, C4, and W, respectively, at 18~GHz.

\subsection{Frequency Spectra}\label{section3.2}

\begin{table}
\tbl{Coordinates of radio sources.}{%
\begin{tabular}{lccc}
\hline
Source ID & Right ascension & Right ascension & Offset \\
& ATCA (J2000.0) & ALMA (J2000.0) & (arcsec) \\
\hline
C1 & 23:44:43.900 & 23:44:43.905 & -0.005 \\
C2 & 23:44:43.857 & 23:44:43.887 & -0.030 \\
C3 & 23:44:43.926 & 23:44:43.973 & -0.047 \\
C4 & 23:44:43.869 & no detection & \\
W & 23:44:41.666 & 23:44:41.661 & 0.005 \\
\hline
Source ID & Declination & Declination & Offset \\
 & ATCA (J2000.0) & ALMA (J2000.0) & (arcsec) \\
\hline
C1 & -42:43:12.508 & -42:43:12.548 & -0.040 \\
C2 & -42:43:10.106 & -42:43:10.684 & -0.578 \\
C3 & -42:43:12.965 & -42:43:13.493 & -0.528 \\
C4 & -42:43:12.131 & no detection & \\
W & -42:43:22.090 & -42:43:22.139 & -0.049 \\
\hline
\end{tabular}}\label{t02}
\end{table}

Frequency spectrum is a clue to understand the radiation mechanism, which will be discussed in Section 4.1. In order to quantify the spectrum, we define the spectral index, $\alpha$, of the flux-density spectrum as $I_\nu \propto \nu^{\alpha}$, where $\nu$ is the frequency. In order to estimate the in-band spectral index, we made four images whose center frequencies were 16488~MHz, 17512~MHz, 18488~MHz, and 19512~MHz each with a 1024 MHz bandwidth. With a simple power-law form, we carried out the error-weighted least-square fit of the four data points, which are shown in figure 2. We obtained the in-band spectral index of $\alpha=-0.50$ for the brightest Source C1. The other Sources showed in-band flux fluctuations due likely to calibration errors and are too faint to derive the in-band spectral index.

In order to obtain wideband spectra of the resolved sources, we combined ALMA Band 3 data of the Phoenix cluster (see \cite{kit20} for details). The center frequency is 92 GHz and the beam FWHM is $1 \farcs 9 \times 1 \farcs 6$ (using $>30k\lambda$ baselines). In the ALMA continuum map, four radio sources are detected with coordinates shown in table~\ref{t02}. We confirm that the coordinates of the brightest and second brightest ALMA sources are identical to the coordinates of Sources C1 and W, respectively, where the statistical position uncertainty of Source C1 is $\sim$(beam size)~/~$\sqrt{S/N} \sim 1''/\sqrt{365} \sim 0\farcs05$ for ATCA ($0\farcs004$ for ALMA), and that of Source W is $0\farcs11$ for ATCA ( $0\farcs05$ for ALMA). Therefore, we conclude that they are cross-matched and the coordinate systems of the two observations are identical at a $0\farcs05$ level. Based on this fact, the declination offset of Source C2 between ATCA and ALMA is less than $3\sigma$ level; the position uncertainty of Source C2 is $0\farcs2$ for ATCA ($0\farcs1$ for ALMA). We also confirm that the position of Source C3 matches between ATCA and ALMA within similar uncertainty.

\begin{figure}[tp]
\begin{center}
\FigureFile(70mm,70mm){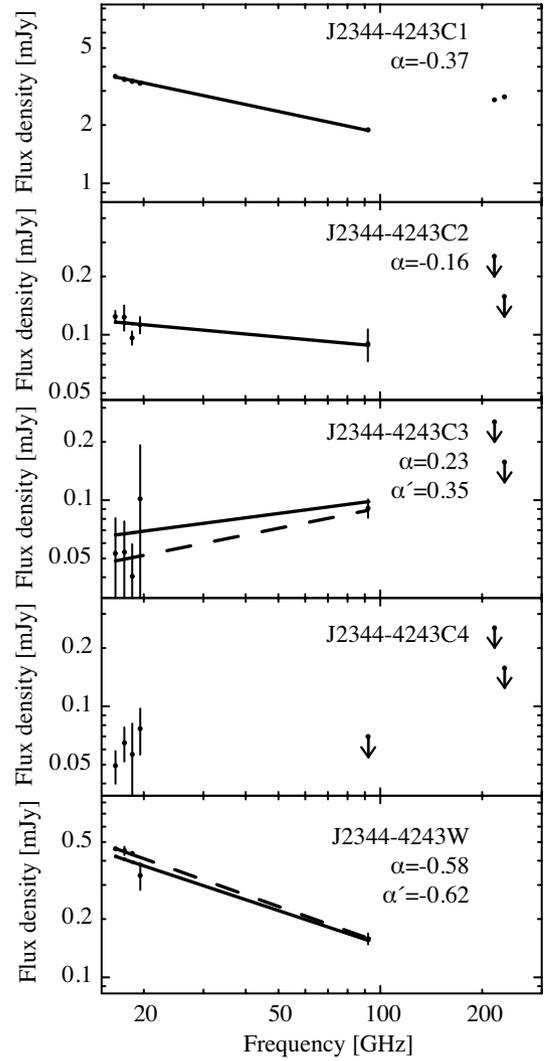}
\end{center}
\caption{
Frequency spectra of Resolved Sources. ATCA 18 GHz, ALMA 92 GHz, 217.7 GHz, and 232.6 GHz data are shown. Arrows are the upper limits of ALMA 217.7 GHz and 232.6 GHz. The solid lines show the best power-law fits and those spectral indices, $\alpha$, are indicated. Also shown as the dashed lines and $\alpha'$ are the best fits and the spectral indices, respectively, except the data at 19.5 GHz.
}
\label{f02}
\end{figure}

The results of the spectral index measurement using ATCA 18 GHz and ALMA 92 GHz data are shown in figure~\ref{f02}. We obtain a flatter ($\alpha=-0.37$) spectrum for Source C1 compared to the in-band spectrum (with $\alpha=-0.50$). Source C2 has a flat spectrum with $\alpha=-0.16$. We also plot ALMA Band 6 \citep{rus17} upper limits to confirm the consistency of the flat spectrum, though we do not obtain strong constraints from them. For Source C3, we obtain a positive slope, $\alpha=0.23$ and $0.35$, with and without the data at 19.5 GHz in the fit, respectively, where there is a large flux error at 19.5 GHz. The positive slope is consistent with the upper limits of ALMA Band 6. At a $5\sigma$ upper limit (or 0.07 mJy) at 92 GHz, the spectrum of Source C4 could be flat. Otherwise, it has a negative slope. Finally, for Source W, $\alpha= -0.58$ and $-0.62$ with and without the data at 19.5 GHz in the fit, respectively.

\section{Discussion}
\label{section4}

\subsection{Possible Radiation Mechanisms}
\label{section4.1}

Synchrotron radiation from cosmic-ray electrons is predominant at centimeter wavelengths in general. Such cosmic rays can be made by AGN and shock waves associated with jets. Typical spectral indices of a radio lobe and an AGN core (or a quasar) are $\alpha \sim -0.7$ and $\alpha \sim 0$, respectively (see e.g., \cite{far14}). The latter flat spectrum is thought to be produced by the synchrotron self-absorption of dense circum-nuclear medium known as a Gigaheltz Peaked Source (GPS), which is often seen in BCGs (Hogan et al. 2015ab). In addition, recent VLBI observations of the 3C84 jet suggest strong synchrotron absorption and a flip of the motion \citep{nag17, hiu18, kin18}. They interpreted that these features are caused by molecular clumps with the typical number density of $10^{3-5}$~${\rm cm^{-3}}$. Therefore, a flat synchrotron spectrum may also be produced by dense molecular clumps. Note that in a case of GPS-like component the spectrum has a peak at a frequency lower than that of our observations, and the spectrum has a positive gradient in an even lower frequency because of the frequency dependence of the absorption.

Free-free emission from thermal electrons can contribute to the radiation at centimeter wavelengths. Free-free emission has a flat spectrum and can be more significant than synchrotron radiation at higher frequencies (see e.g., \cite{con92} for a review). Thermal electrons emitting free-free emission exist in warm ionized medium of the interstellar medium (ISM) ionized by radiation from massive stars and AGN, and supernova shocks. Actually, \citet{mcd14} suggested that the low-ionization lines from a complex emission-line nebula at the BCG are consistent with photo-ionization by young stars. Another origin of warm ionized medium may be a transient warm-ionized ICM of the cooling flow at the cluster center.

The spinning dust emission \citep{dra98a, dra98b} can also contribute to the radiation at centimeter wavelengths, if a massive amount of small dust grains exists. The emission can have a flat or positive spectral index around the relevant frequency in general. The brightness of the spinning dust emission is estimated to be  $O(0.01)(n_{\rm H}/10^2~{\rm cm^{-3}})$~mJy/beam around 15--30 GHz, where $n_{\rm H}$ is the hydrogen number density. Even higher frequency, galactic thermal dust emission becomes predominant, showing a steep, positive slope. The spinning dust emission should spatially correlate with thermal dust emission.

In the following subsections, we discuss the origins of the observed radio sources near the cluster center. We focus on the compact emissions at the BCG, the extended emission around the BCG, and the diffuse emission extending to the north and south directions in this order.

\subsection{AGN Core and Jets (Sources C1, C3, and C4)}
\label{section4.2}

\begin{figure}[tp]
\begin{center}
\FigureFile(80mm,80mm){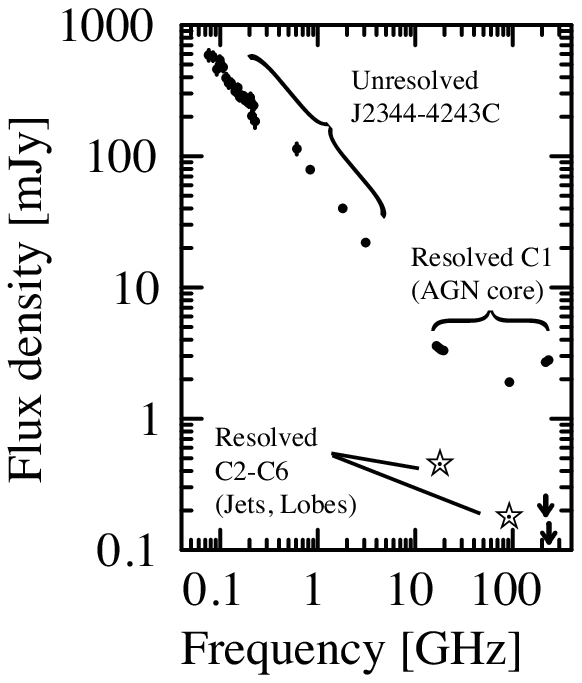}
\end{center}
\caption{
Broadband radio spectrum of J2344-4243C. The integrated flux densities are taken from MWA GLEAM (76--227 MHz) \citep{hur17}, GMRT (610 MHz) \citep{van14}, SUMSS (843 MHz) \citep{boc99}, and ATCA (1.8 GHz and 3.1 GHz). These observations did not resolve the substructures (C2--C6). Also shown are this ATCA work (18 GHz), ALMA Band 3 (92 GHz) \citep{kit20}, and ALMA Band 6 (218 GHz and 233 GHz) \citep{rus17}. For those three data, the flux densities of Source C1 are shown as black dots. Open stars show the sum of the flux densities for C2, C3, C4, C5, and C6 at 18 GHz and the sum of C2 and C3 at 92 GHz. Arrows indicate the upper limits at 218 GHz and 233 GHz.
}
\label{f03}
\end{figure}

Source C1 is a compact source and is located at (R.A., Dec.)  = (23:44:43.90, -42:43:12.51) (J2000) with the positional accuracy $\sim 0\farcs05$. We confirmed that the coordinate is identical to the X-ray brightness center reported by \citet{mcd15} (R.A., Dec.) = (23:44:43.94, -42:43:12.52) (J2000), so that Source C1 is likely the AGN core. \citet{mcd14} reported the coordinate of the optical brightness center of the BCG as (R.A., Dec.) = (23:44:43.96, -42:43:12.20) (J2000). The position of the BCG center can be identical to that of Source C1, according to the fact that there is the astrometry offset between HST and ATCA/ALMA, ($\delta$R.A., $\delta$Dec.) = (-0\farcs08, +0\farcs3) \citep{dun17} which is very close to the offset we found.

Sources C3 and C4 are located near the AGN core and they are along the north-south diffuse emission with straight bar-shape structures. These structures bring an idea that Sources C3 and C4 are bipolar jets from the AGN, where flat or positive spectra of them could be explained by self-absorbed synchrotron radiation. It is interesting that the spectral index of Source C3 is different from that of Source C4, although they are located at the point-symmetric positions with respect to Source C1. Thus, there should exist inhomogeneous structure around the AGN. 

Figure~\ref{f03} shows the broadband radio spectrum of J2344-4243C. We plot the archival data of MWA GLEAM \citep{hur17} at 76--227 MHz, where J2344-4243C is not spatially resolved. The data indicate the spectral index of $\alpha$(GLEAM) $=-0.929$. Such a slope is also found by \citet{van14} who reported the spatially-integrated flux density of 114.0 $\pm$ 11.0 mJy at GMRT 610 MHz and obtained $\alpha$(GMRT) $=-0.84$ for GMRT 156 MHz and 610 MHz. However, \citet{mcd14} reported the steeper spectrum with the spectral index of $\alpha =-1.35$, based on the unresolved data of SUMSS \citep{boc99} 843 MHz (79.2 ± 3.0 mJy) and ATCA 16 cm. A similar steep spectrum of $\alpha =-1.2$ is obtained from the previous ATCA observation (ID: C2585) with the total intensities 40~mJy at 1.8~GHz and 22~mJy at 3.1~GHz. 

While the above steep spectrum has been observed below several GHz, we found a shallower spectrum with the spectral indices of $\alpha$(ATCA) $= -0.50$ and $\alpha$(ATCA + ALMA) $= -0.37$ in the frequency range of $\sim 10$--100~GHz. The broadband spectrum is thus difficult to explain with a single power law. We suspect that the broadband spectrum of Source C1 consists of at least two components. One, which dominates emission at $\sim 10$--100~GHz, can be self-absorbed synchrotron radiation like a GPS source. Actually, the AGN core is highly obscured \citep{ued13}, suggesting significant absorbers around the AGN. The other can be optically-thin synchrotron radiation dominating emission below several~GHz. We show the sum of the flux densities for the other resolved components in figure~\ref{f03}. The sum is about an order-of-magnitude fainter than Source C1, providing an upper limit of the strength of spectrum cut off, i.e. the population of the highest-energy electrons emitting synchrotron radiation.

We also plot the data of ALMA 233 GHz and indicates that the flux density of Source C1 increases as the frequency increases. \citet{mcd14} displayed three SED models for the BCG; elliptical galaxy, Type 2 QSO, and M82-like star-burst galaxy. ALMA 233 GHz is consistent with the Type 2 QSO model (not shown, see \cite{mcd14}), which is also consistent at optical bands. Another possibility is that the GPS component is time variable \citep{hog15b}. In this case, the lower flux in Band 3 (observed in 2016) compared to ATCA 18 GHz and Band 6 (observed in 2014) is due to time variation, and these emissions are explained by a single GPS.

\subsection{Extended Emissions Surrounding the AGN}
\label{section4.3}

\begin{figure}[tp]
\begin{center}
\FigureFile(80mm,80mm){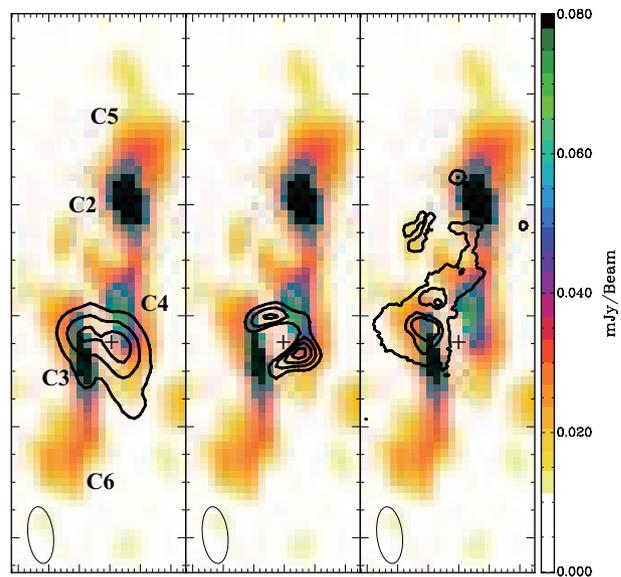}
\end{center}
\caption{
The superposed images of the Phoenix cluster center. The cross indicates the position of Source C1, which is subtracted from the background image. The synthesized beam pattern is shown at bottom-right. (Left) The ALMA CO(3-2) integrated intensity map \citep{rus17}. The contour levels are 0.7, 1.5, 3.1 mJy/beam. (Middle) The ALMA Band 6 (233 GHz) continuum contour map \citep{rus17}, where the primary compact source at the peak of the continuum intensity is subtracted. The contour revels are 0.5, 1.0, 1.5, and 2.0 mJy/beam. (Right) The HST 814 nm contour map. The contour levels are 0.2, 0.7, 1.5 in an arbitrary unit.
}
\label{f04}
\end{figure}

In addition to Sources C1, C3, and C4, our ATCA observation suggests extended emissions which seem to surround the AGN core (see Fig.~\ref{f01}). In order to look for possible counterparts, we made superposed images with ALMA CO(3-2) and ALMA 233 GHz continuum \citep{rus17}, and {\it Hubble Space Telescope} ({\it HST}) F814W (814 nm, I band, \cite{mcd15}, the coordinate offset is not corrected) in figure~\ref{f04}. Here, the CO(3-2) emission indicates star-forming molecular gas. The 233 GHz continuum represents mostly the dust distribution. The 814 nm emission is a tracer of stars and star formation. Star formation implies UV radiation field of OB stars and their supernovae, i.e. the existence of the warm ionized gas and cosmic-ray electrons. The image clearly indicates that there are CO(3-2), 233 GHz continuum, and/or 814 nm emission around Source C1, suggesting the presence of a large amount of molecular gas, dust grains, and non-thermal electrons, all of which can contribute radio emission at 18 GHz (28.7 GHz at $z=0.596$). Therefore, the extended emission surrounding the AGN could be superposition of jets, ISM, and circum-galactic medium.

In the northwestern part of the central extended emission including Source C4, the 233 GHz emission (dust) is bright. Therefore, radio emission (18 GHz and 92 GHz) of Source C4 could be superposition of jet synchrotron and dust emission. It should be noted that the 814 nm emission (stars) is not bright and the difference between dust emission and 814 nm emission can not be explained by relative coordinate errors (by $0\farcs3$) between the radio and optical observations. One possible explanation of the difference is that there is an inclined BCG and its gas disk obscures star lights from the southwestern region, i.e. dust lane. This potential galactic disk is partly seen as the high-density molecular gas of CO(3-2). Actually, \citet{rus17} claimed the east-west velocity gradient which is consistent with ordered motion or rotation about the AGN, although we need a higher spatial resolution CO observation in order to confirm the existence of such a putative disk. If the dust lane is the case, the radio jet is edge-on and rather parallel to the disk.

In the southeastern part of the central extended emission including Source C3, the 814 nm emission (stars) is bright but the 233 GHz continuum emission (dust) is not. Therefore, the radio emission (18 GHz and 92 GHz) of Source C3 could be superposition of jet synchrotron and free-free emission from the warm gas ionized by UV radiation from massive stars. Meanwhile, the spinning dust emission could be minor, because a feature-less, flatter extinction curve implies that the size distribution of the dust is skewed toward larger grains \citep{mcd15}, i.e. the spinning dust emission from small dust grains may be less important compared to the typical case \citep{dra98a, dra98b}.

\subsection{Radio Lobes (Sources C2, C5, and C6)}
\label{section4.4}

\begin{figure}[tp]
\begin{center}
\FigureFile(80mm,80mm){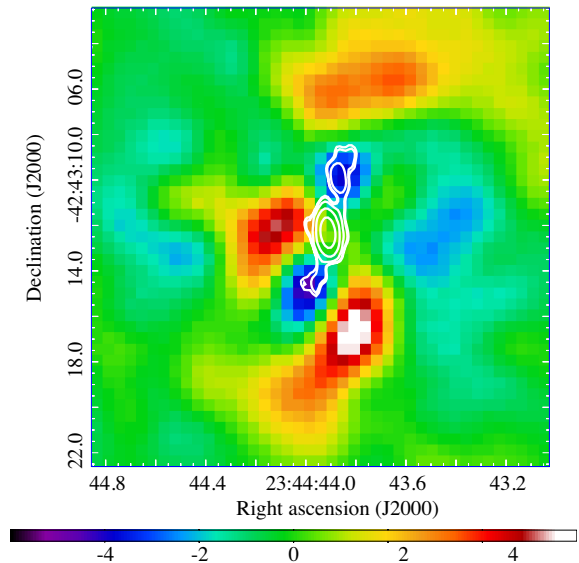}
\end{center}
\caption{
The superposed image of the Phoenix cluster. The background color shows the {\it Chandra} X-ray surface brightness residual in an arbitrary unit. The white contours are ATCA 18 GHz emission of 0.02, 0.025, 0.05, 0.2, and 1 mJy/beam.
}
\label{f05}
\end{figure}

We examine ICM distribution around the AGN and compare it with the 18 GHz radio distribution. For this purpose, we made an X-ray surface brightness residual map using {\it Chandra X-ray Observatory} archival data of the Phoenix cluster  (ObsID: 13401, 16135, and 16545). The energy range of soft X-rays, $0.5 - 2.0$\,keV, is adopted to reduce the contamination emission from the type 2 QSO. With the same dataset and energy range, \citet{mcd15} already showed the residual map, but we reprocessed the data using {\it Chandra} Interactive Analysis of Observations (CIAO; \cite{fru06}) version 4.9 and new calibration database (CALDB) version 4.7.8. \citet{mcd15} modeled the X-ray surface brightness with a two-beta model. We modeled the X-ray surface profile in $1'' < r < 5''$ from the position of the AGN in X-rays using an ellipse model \citep{ued17, ued18}, where the position angle is $-86$~degree and the axis ratio of the obtained ellipse is 0.92. We calculated the mean profile of the X-ray surface brightness using the parameter of the obtained ellipse and subtracted it from the original X-ray surface brightness map.

The resultant X-ray residual map superposed with the 18~GHz radio image is shown in figure~\ref{f05}. The residual structure is in broad agreement with that shown by \citet{mcd15}. Interestingly, both the Northern and Southern structures are aligned with X-ray negative regions (i.e. cavities). This supports that Sources C5 and C6 are radio lobes, as often seen in the nearby Universe. The northern extended structure of HST 814 nm broadly aligns with the northern bar structure of the 18 GHz emission, suggesting that a part of star-forming gas is blown away from the BCG by the jet.

We estimate non-thermal energy (cosmic-rays and magnetic field) associated with the diffuse radio emission under the assumption of energy equipartition (see \cite{aka18b}). Our detection of 0.03 mJy/beam at 18 GHz for Sources C5 and C6 and representative values ($p$, $\gamma_{\rm min}$, $\gamma_{\rm max}$, $D_{\rm los}$) = (3, 200, 3000, 10~kpc) gives $B\sim 23.4 f^{1/4}$~${\rm \mu G}$ or the cosmic-ray electron density, $n_{\rm ce} \sim 7.10 \times 10^{-8} f^{-1/2}$~${\rm cm^{-3}}$. Here, $p$ is the index of cosmic-ray electron energy distribution ($p=3$ corresponds to $\alpha=-1$ in this paper), $\gamma_{\rm min}$ and $\gamma_{\rm max}$ are the minimum and maximum Lorentz factor of the cosmic-ray electrons, respectively,  and $D_{\rm los}$ is the depth of the emission along the LOS. $f-1$ is the energy fraction of CR-ions with respect to CR-electrons (see, e.g., \cite{pfr04, aka18b}). Although the value of $f$ is unclear, it is $O(100)$ for radio halos in Perseus and Coma clusters \citep{pfr04}. The resultant magnetic-field strength and the cosmic-ray electron density are much larger than those known in nearby cluster environments, supporting that these diffuse emission is radio lobes.

We have checked that there is no clear optical counterpart of Source C2. Since it is a compact source, one may consider that Source C2 is an unresolved radio galaxy or a quasar. However, supposing the population of 0.1 mJy sources at 18 GHz  (e.g. \cite{wil08}), the probability that an extragalactic radio source overlaps a 1 arcsec$^2$ patch is $1.2\times 10^{-3}$~\%, which is very rare. Therefore, a reasonable interpretation is that Source C2 is an object associated with the diffuse emission and it is likely a bright knot and/or kinks of jets seen in, for example, the BCGs of Abell 1795 and Abell 2597. In other words, Sources C2, C5, and C6 are an older FRII and Sources C3 and C4 are a younger FR I of another two-side jet. Jet precession may explain the fact that the axes of C2-C1 and C4-C1-C3 are not exactly aligned each other.

The diffuse sources C5 and C6 are located $\sim 13$~kpc and $\sim 20$~kpc away from Source C1, respectively. Assuming that the shock waves caused by jets are propagating with the sound velocity of 900~km s$^{-1}$ (3keV, \cite{ued13}), we obtain 12~Myr and 18~Myr for the diffuse sources C5 and C6, respectively. Moreover, if C3 and C4 are newly-born jets, their ages are $\sim 1$~Myr. Therefore, if Sources C5 and C6 are radio lobes of old jets and C3 and C4 are newly-born jets, the duty cycle of AGN jet is $O(10)$~Myr. It is interesting that the new jets may link with young ($\sim$Myr, \cite{mcd15}) stellar population; launches of the new jets and massive star formation happened at the nearly same time. Our observation did not find even older radio jets or lobes. If there were multiple jet launches in the past, with future high sensitivity and high angular-resolution X-ray observations we could see the ripples of X-ray cavities as seen in the Perseus cluster \citep{fab06}.

We estimate the rest-frame, monochromatic radio power at the rest-frame frequency $\nu'$ as
\begin{equation}
L_{\rm \nu'} = 4\pi D_{\rm L}^2 \Omega (1+z)^{-1} S_\nu\left\{\frac{\nu'}{(1+z)}\right\}^\beta,
\end{equation}
where $D_{\rm L}$ is the luminosity distance, $\Omega$ is the emission area, $z$ is the redshift, $S_\nu$ is the observed surface brightness at the observed frequency $\nu$, and $\beta$ is the spectral index for $S_\nu \propto \nu^{\beta}$. We adopt $D_{\rm L}\sim 3.53~{\rm Gpc} \sim 1.089\times 10^{26}~{\rm m}$, $\Omega=1$ arcsec$^2$ for Source C5 or C6, $z=0.596$, and $S_\nu = 0.03$~mJy/beam at 18~GHz with the beam size $0.537$~arcsec$^2$ with $r=0.5$. Assuming the spectral indices of $\beta=-0.5$ and $-1.0$, the monochromatic radio powers at $\nu' =1.4$ GHz are $L_{\rm 1.4}=2.4 \times 10^{23}$~W/Hz and $1.1 \times 10^{24}$~W/Hz, respectively. This radio power is close to the 50\% percentile of the cumulative luminosity function for BCGs \citep{hog15a, hog15b}.

We then estimate the radio luminosity, 
\begin{equation}
L = \int_{\nu'_1}^{\nu'_2} L_{\rm \nu'} d\nu'.
\end{equation}Integrating emission from $\nu'_1=10$~MHz to $\nu'_2=10$~GHz for $\beta=-0.5$ and $-1.0$, we obtain $L \sim 8.6\times 10^{39}$~erg/s and $2.1\times 10^{40}$~erg/s, respectively. The derived radio power of the diffuse emission is not enough at all to compete with the X-ray radiative cooling loss,  $L_{\rm cool,100kpc}$ = $9.6 \times  10^{45}$ erg/s \citep{mcd15}. Even if we integrate the whole northern emission including Sources C2, C4, C5, and a bar-shape structure, the radio luminosity increases only by a factor of 4. The derived radio power is also small compared to the jet power according to X-ray cavities, $L_{\rm cavity}$ = $2-7 \times  10^{45}$ erg/s \citep{mcd15}, implying that the jet energy is dominated by kinetic energy.

\section{Concluding Remarks}
\label{section6}

SPT-CL J2344-4243 (the Phoenix cluster) located at the redshift of 0.596 has an extreme star-burst BCG at the cluster center. The BCG possesses a heavy SMBH surrounding a huge amount of cold gas. These features suggested the realization of the classical picture of cooling flow in the previous works. Previous X-ray observations indicated cool gas in the cluster core. They also showed X-ray cavities, suggesting the existence of AGN feedback often seen in nearby cool-core clusters. However, radio jets/lobes of AGN have never been spatially resolved previously.

We conducted the ATCA 15 mm observation of the Phoenix cluster center, and resolved emissions at/around AGN and those extending toward the north-south direction. At the cluster center, we find a compact radio source associated with the AGN. We also find two compact sources near the AGN and they are likely newly-born jets. Moreover, surrounding the AGN, we find the extended emission which could primary be synchrotron radiation partially absorbed by dense molecular clumps and free-free emission from the massive warm ionized gas. Because of the difference between the potential dust distribution and the extended emission, the spinning dust emission from dusty circum-galactic medium could be secondary.

We discovered diffuse bipolar radio structures extending from the AGN of the BCG toward the north-south direction. The locations of the radio structures are consistent with X-ray cavities, implying that they are radio jets/lobes of the AGN. We find a bright jet knot and/or kink candidate at the north. The lobe sizes are only 10--20 kpc and the ages could be less than a few tens of Myr. Moreover, if compact radio emissions near the center is newly-born jets, those ages are $\sim 1$~Myr. We point out that launches of new jets and massive star formation with young ($\sim$Myr, \cite{mcd15}) stellar population happened at the nearly same time in the past. Therefore, there may exist the same trigger which caused both activities. The monochromatic radio power of a radio lobe is estimated to be $O(10^{23})$ W/Hz at 1.4 GHz. We stress that this radio power is not enough to recover the X-ray cooling loss of the ICM. 

Our study indicates that a sub-arcsec scale angular resolution is required to study the radio emission at the center of the Phoenix cluster. Since the radio lobes should be synchrotron dominated, a lower frequency observation has advantages in the signal-to-noise ratio. Naively, an angular resolution at 18~GHz with a 6~km baseline (this work) is equivalent to that at 0.9 GHz with a 120~km baseline, which will be indeed available with the phase 1 of the Square Kilometre Array (SKA)\footnote{https://www.skatelescope.org}. Therefore, this cluster can be a good use case for the SKA.

\vskip 12pt
We thank the referees for their useful comments and suggestions. The Australia Telescope Compact Array is funded by the Commonwealth of Australia for operation as a National Facility managed by CSIRO. The scientific results of this paper are based in part on data obtained from the Chandra Data Archive: ObsID 13401, 16135, and 16545. This work was supported in part by JSPS KAKENHI Grant Numbers, JP15H03639(TA), JP15H05892 (MO), JP15K17614(TA), JP17H01110(TA), JP17H06130 (KK, RK), JP18K03693 (MO), JP18K03704 (TK), and by the Ministry of Science and Technology of Taiwan (grant MOST 106-2628-M-001-003-MY3) and by Academia Sinica (grant AS-IA-107-M01). 

\vskip 12pt
{\it Note added in proof: After this paper was submitted, \citet{mcd19} submitted a paper and presented a radio continuum image in the 8--12 GHz X-band using the Karl G. Jansky Very Large Array (VLA). \citet{mcd19} found multiple radio sources. While the synthesized beam of the JVLA data is quite elongated, the brightest radio source of \citet{mcd19} seems to associate with C1+C3+C4, and the northern radio source of \citet{mcd19} does C2+C5.}

\end{document}